\documentclass[conference,a4paper]{IEEEtran}
\IEEEoverridecommandlockouts

\normalsize
\usepackage{cite}
\usepackage[normalem]{ulem}
\usepackage{color}
\ifCLASSINFOpdf
   \usepackage[pdftex]{graphicx}
   \DeclareGraphicsExtensions{.pdf}
\else
   \usepackage[dvips]{graphicx}
   \DeclareGraphicsExtensions{.eps}
\fi

\usepackage{amsmath,amssymb,amsfonts,mathtools}
\usepackage[cmintegrals]{newtxmath}
\usepackage{bm}
\usepackage[tight,footnotesize]{subfigure} 

 \usepackage{url}

{}

\DeclareMathOperator{\diag}{diag}

\usepackage{paralist}
\usepackage{enumerate}
\usepackage{array}

\usepackage{textcomp}
\usepackage{xcolor}
\def\BibTeX{{\rm B\kern-.05em{\sc i\kern-.025em b}\kern-.08em
    T\kern-.1667em\lower.7ex\hbox{E}\kern-.125emX}}

\begin{document}
%
\title{Hybrid Beamformer Design for High Dynamic Range Ambient Backscatter Receivers}%

\author{\IEEEauthorblockN{
Ruifeng~Duan\IEEEauthorrefmark{1},
Estifanos~Menta\IEEEauthorrefmark{1},
H{\"u}seyin~Yi\u{g}itler\IEEEauthorrefmark{1},
Riku~J\"{a}ntti\IEEEauthorrefmark{1}, and 
Zhu Han\IEEEauthorrefmark{2}}
\IEEEauthorblockA{\IEEEauthorrefmark{1}Department of Communications and Networking, Aalto University, 02150 Espoo, Finland\\
Email: \{ruifeng.duan; estifanos.menta; yusein.ali; riku.jantti\}@aalto.fi}
\IEEEauthorblockA{\IEEEauthorrefmark{2}Department of Electrical and Computer Engineering, University of Houston, Houston, TX 77004, USA\\
Email: zhan2@uh.edu}
}


\maketitle

\begin{abstract}
In bi-static Ambient Backscatter Communications (AmBC) systems, the receiver needs to operate at a large dynamic range because the direct path from the ambient source to the receiver can be several orders of magnitude stronger than the scattered path modulated by the AmBC device. In this paper, we propose a novel analog-digital hybrid null-steering beamformer which allows the backscatter receiver to detect and decode the weak AmBC-modulated signal buried in the strong direct path signals and the noise without requiring the instantaneous channel state information. The analog cancellation of the strong signal components allows the receiver automatic gain control to adjust to the level of the weak AmBC signals. This hence allows common analog-to-digital converters to be used for sampling the signal. After cancelling the strong components, the ambient source signal appears as zero mean fast fading from the AmBC system point of view. We use the direct path signal component to track the phase of the unknown ambient signal. In order to avoid channel estimation, we propose AmBC to use orthogonal channelization codes. The results show that the design allows the AmBC receiver to detect the backscatter binary phase shift keying signals without decoding the ambient signals and requiring knowledge of the instantaneous channel state information.


\end{abstract}


\section{Introduction}

Ambient backscatter communication (AmBC) introduced in \cite{Liu2013}, one of the recent breakthrough technologies selected by MIT Technology Review in 2016, has been emerging and captured much attention. Without need of a dedicated power infrastructure and a carrier emitter, ambient backscatter communication (AmBC) enables devices to communicate by backscattering ambient modulated radio frequency (RF) signals existing in the atmosphere, such as TV broadcasts \cite{Parks2013, Parks2014a}, cellular communication systems \cite{Parks2013, Shen2017}, frequency modulation (FM) broadcasts \cite{Wang2017, Yang2017}, Bluetooth \cite{Iyer2016}, and WiFi transmissions \cite{Kellogg2016, Zhang2016, Zhang2016b}. Hence, AmBC has a power efficiency which is orders of magnitude higher than traditional radio communications \cite{Liu2013}. Since then researchers have developed numerous conceptually proved applications of AmBC technologies in: 1) medical science \cite{Iyer2016}; 2) environmental monitoring \cite{Daskalakis2017a, Naderiparizi2017}; 3) network communications \cite{Liu2013,Iyer2016, Hoang2017a}. In addition, AmBC improves spectral efficiency because an AmBC link can co-exist with a legacy wireless system under proper system design as shown in recent work \cite{Duan2017, Darsena2017,Liu2018}.

\subsection{State of the Art}
In addition to using commodity receivers, various receiver designs for backscatter communications have been proposed and studied in literature. Based on a time-varying change of the wireless channel caused by the on-off keying (OOK) modulation, the receiver in \cite{Kellogg2014} exploits the frequency diversity of the wideband WiFi signals, and then conducts a selective and linear combination of information across the good sub-channels. The work of \cite{Yang2017} adopts universal wideband backscatter reader exploiting maximal ratio combining with carrier interference cancellation techniques. Besides the practical receiver designs, a large number of signal detection methods for backscatter receivers have been considered in theoretical analysis, such as energy detection \cite{Liu2017d}, maximum likelihood (ML) detection, covariance matrix (CM)-based detection, and maximum a posteriori (MAP) detection. To improve the bit error rate (BER) at low signal-to-noise ratio (SNR), a CM-based signal detection algorithm was considered in \cite{Zeng2016} for signal-antenna AmBC systems using the ratio of the two CMs representing that the backscatter signal is absent or present. Authors in \cite{Wang2016} considered using a joint energy and ML detector at the receiver to detect the backscatter information bits. The receiver developed in \cite{Lee2017} is able to determine the signal detection thresholds for WiFi backscatter systems with a multi-antenna receiver based on the maximum distance of the received power (pilot sequences) at each antenna. Assuming that the channel status information (CSI) is available at the receiver, \cite{Liu2017b} proposed to use an MAP detector to decode the backscatter information bits. 

Most recently, authors in \cite{Darsena2017, Yang2018a, Duan2018a, ElMossallamy2018} considered the ambient orthogonal frequency-division multiplexing (OFDM) modulated signals transmitted by the legacy systems. In \cite{Darsena2017}, the ML rule was applied to obtain the ML decision regions to minimize the BER. Exploiting the repeating structure of the ambient OFDM signals and the test statistics, \cite{Yang2018a} has proposed a receiver design to cancel out the direct-link interference generated by the legacy transmitter. In \cite{ElMossallamy2018}, a noncoherent  backscatter  communication technique  over ambient OFDM  signals has been proposed by using null subcarriers. In addition, the successive cancellation technique can be applied to jointly decode the legacy system and backscatter system information for a co-located legacy and backscatter receiver \cite{Duan2017}. As a large number of OFDM subcarriers contain repetitive elements inducing time correlation, \cite{Duan2018a} proposed a simple receiver for a Binary Phase Shift Keying (BPSK)-modulated AmBC system without knowledge of CSI, the statistical channel covariance matrices, and the noise variance at the receiver antennas.

\subsection{Challenges}\label{sec:challenges}
The communication performance of AmBC systems depend on the signal-to-interference-plus-noise ratio (SINR) of the scattered ambient signal, on top of which the backscatter information is modulated. The analog front-end of a receiver of a contemporary communication system is usually configured to utilize the dynamic range of the analog-to-digital converter (ADC) to capture as much information (in terms of bits) as possible on the desired signal by pushing the undesired components and noise sources (including the quantization noise) to a few least significant bits. For the ambient backscattering systems, however, this strong signal component of the legacy system (source of the ambient signal) drastically degrades the SINR of the AmBC signal\footnote{In this paper, we assume that the designed simple backscatter receiver cannot decode the ambient signal.}. The AmBC signal is restrained to a few least significant bits of the ADC output resulting in a extremely low SINR for the AmBC. Therefore, to improve the SINR of the AmBC, the receiver needs to first eliminate the strong direct-path ambient signal, and then decode the AmBC information. Regardless of the receiver architecture, after the first step, the achieved SINR is still low due to the high quantization noise\footnote{Here, high quantization noise refers to spectral density ratio of the ambient signal to the spectral density of the quantization noise. Since the digital filtering cannot eliminate the in-band noise, large quantization noise cannot be eliminated using conventional filters.} unless a very high dynamic range (and high cost) ADC is used. The high power difference between the direct and scattered signal paths also make angle-of-arrival (AoA) estimation and receiver beamforming challenging. AoA estimation can be easily done for the strong direct path component, but simultaneously finding the weak scattered signal component is a demanding task. 

The ambient signal has fast varying random phase making the use of coherent modulation schemes such as non-return-to-zero (NRZ) binary phase shift keying (BPSK) challenging at the AmBC device. If the channel phases are slowly varying, the differential phase shift keying (DPSK) modulation schemes can be utilized. In AmBC, since a legacy system has much higher data transmission rate than the AmBC system, the phase of the legacy system symbol varies rapidly from one to another resulting in very fast fading seen by the AmBC link. Consequently, many AmBC modulation schemes use non-coherent modulation such as on-off keying (OOK) \cite{Ishizaki2011} or Frequency Shift Keying (FSK) \cite{Ensworth2015}. Alternatively, they combine coherent modulation with frequency modulation \cite{Vougioukas2016}. FSK and frequency-shifted BPSK require that the adjacent bands of the ambient signal are empty. If the phase of the ambient signal can be known at the AmBC receiver, coherent modulation schemes can be utilized directly without having to add frequency modulation leading to better spectral efficiency. However, this in general is not practical.

\subsection{Contributions and Paper Organization}
To address these issues, we develop a two-stage hybrid analog-digital null-steering beamformer for a high dynamic range ambient backscatter receiver. The analog part is first utilized to separate the direct and scattered paths, and then the digital processing is applied to form a beam towards the desired signal. By first separating the strong direct and weak signal components at the analog part, automatic gain control and ADC can be applied to both components separately. This design allows the receiver to deal with the high power differences between the signal components.

We use the received direct path component to track the unknown phase of the ambient signal by correlating the direct path component with the scattered path signal. The phase of the corrector output signal only depends on the channel phase and the AmBC modulator symbol phase. To avoid the need to estimate the channel, we consider that the AmBC modulator uses orthogonal Hadamard channelization codes to represent the different AmBC signals. The second correlator is then utilized to demodulate the AmBC symbols.  The performance of such design is evaluated via simulations with respect to the AmBC BER. These results confirm that our proposed scheme allows the AmBC device operating at a reasonable distance from the receiver with a practically acceptable BER of $10^{-3}$ without employing error correction techniques. 

The rest of this paper is organized as follows. In Section \ref{sec:system}, we outline the system and signal models. Section \ref{sec:RxOperation} presents our receiver design consisting of a two-stage beamformer and a two-stage correlator. Using this design, we evaluate the BER of the backscatter system in Section~\ref{sec:Sim}. Section \ref{sec:Con} concludes this paper and discusses the future work.

\begin{figure}[!t]
\centering
	\includegraphics[width=0.9\columnwidth]{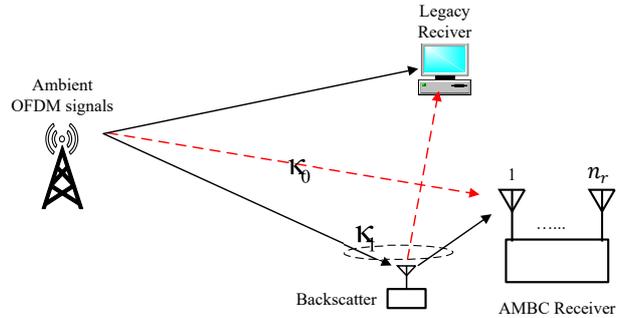}
	\caption{An illustration of the considered system.}
    \label{fig:system}
\end{figure}

\section{System Model}\label{sec:system}
We consider a system shown in Fig.~\ref{fig:system} consisting of two sub-systems: 1) a legacy wireless system transmitter and 2) an AmBC system. The interference to the legacy receiver in general can be ignored in our considered setup\cite{Ruttik2018}. We focus on the AmBC receiver design. The legacy transmitter generates continuous carrier modulation scheme with carrier wavelength $\lambda$. The AmBC system consists of an ambient backscatter device and an $n_r$-antenna receiver forming a linear array with antenna spacing $\Delta d$. Let $s\in\mathbb{C}$ denote the equivalent low-pass representation of the ambient signal generated by the legacy transmitter. The symbol $\mathcal{K}_0$ denotes the set of direct signal paths directly from the ambient transmitter to the receiver that do not pass through the AmBC device. $\mathcal{K}_1$ denotes the set of scattered signal paths that pass through the AmBC device and are being modulated by it. The AmBC device modulates the signal impinging at its antenna by controlling its complex reflection coefficient.

We consider an AmBC device adopting the BPSK modulation scheme, i.e., the information symbol $x\in \{-1, +1\}$. The AmBC device consists of an encoder to spread the AmBC symbols by using orthogonal codewords. In this paper, we apply the Hadamard codes to spread the data symbol so that the BPSK backscatter needs to switch its status in a high rate.\footnote{This practically reasonable spreading scheme can be readily implemented at an ambient backscatter device because an off-the-shelf dialect component can complete the status switching as short as $4 n$s,  e.g., 1N4148 diode \cite{Nexperia2002}.}

We assume that deterministic channels with strong dominant light-of-sight signal paths between the ambient RF source and the AmBC receiver, between the the ambient RF source and the backscatter device, and between the backscatter device and the AmBC receiver. We consider a widely-used path loss model consisting of both frequency-dependent and distance-dependent losses, which is presented as $\mathsf{PL} = (\lambda/(4\pi d))^2$ with propagation distance $d$. Notation $d_0$ denotes the distance between the legacy transmitter and the AmBC receiver, $d_1$ is the distance between the backscatter device and the AmBC receiver, and $d_2$ represents the distance between the legacy transmitter and the backscatter device. The corresponding phase rotation of the RF propagation for each path is considered as $\exp(-j2\pi d_l/\lambda), \; l = {0, 1, 2}$, where $j=\sqrt{-1}$. For instance, the only frequency-dependent loss of the backscattering path can be approximately $-27$dB lower at 500MHz frequency band than that of the direct path. This significant radio propagation characteristic has been ignored in literature. We have highlighted in Section~\ref{sec:challenges} that this phenomenon results in numerous challenges on a practical AmBC receiver design.

We denote the AoA for the $k$-th signal path as $\phi_k$, and define the directional cosine as $\Omega_k=2\pi \Delta d\cos(\phi_k)$. The unit spatial signature at the receiver antenna array in $\Omega_k$ becomes $\boldsymbol{a}(\Omega_k)=\begin{bmatrix} 1 & e^{i\Omega_k} & e^{i2\Omega_k} & \cdots & e^{i(N-1)\Omega_k} \end{bmatrix}^\dagger$, where the superscript ${}^\dagger$ denotes the conjugate transpose of the vector. The baseband channel gains of the direct paths $\boldsymbol{g}_0$ and the backscattering ones $\boldsymbol{g}_1$ are therefore given by \cite{Griffin2010}:
\begin{equation}
    \boldsymbol{g}_0 = \left(\frac{\lambda}{4\pi}\right)^2\frac{e^{-j2\pi d_0/\lambda}}{d_0^2}, \quad \boldsymbol{g}_1 = \left(\frac{\lambda}{4\pi}\right)^4\frac{e^{-j2\pi (d_1+d_2)/\lambda}}{d_1^2 d_2^2},
    \label{eq:pathloss}
\end{equation}
where we assume the same attenuation of the (direct or backscattering) paths for all antenna pairs. The corresponding $m$-th derivative of $\boldsymbol{a}(\Omega_k)$ reads $\boldsymbol{a}^{(m)}(\Omega_k)=\frac{\mathsf{d}^m}{\mathsf{d}\Omega_k^m}\boldsymbol{a}(\Omega_k)$. 
Since that a backscatter information signal is encoded using the Hadamard codes, a length-$M=2^K$ Hadamard codeword with order of $K$ is obtained, e.g., by selecting a row, from a Hadamard matrix of $M\times M$ over $\{+1, -1\}$. Without loss of generality, we select two rows to encode $+1$ and $-1$ of the BPSK symbols generated by the backscatter, respectively. The two encoded sequences over $\{+1, -1\}$ are orthogonal. Define matrices $\boldsymbol{A}_l^{(m)}=\diag[\boldsymbol{a}^{(m)}(\Omega_k),k\in \mathcal{K}_l]$, $l=0,1$. Within one length-$M$ codeword, the received signal vector reads
\begin{equation} \label{eq:y_coded}
    \boldsymbol{y}[i] = \sqrt{\mathsf{snr}}\left(\boldsymbol{A}^{(0)}_0\boldsymbol{g}_0+\boldsymbol{A}^{(0)}_1\boldsymbol{g}_1\tilde{x}[i]\right)s[i]+\boldsymbol{z}[i], \quad \forall i=1,\cdots, M,
\end{equation}
where $\mathsf{snr}$ represents the received SNR of the ambient signal source (legacy transmitter), $\tilde{x}[i],  \forall i=1,\cdots, M$ denotes the $i$-th element of an encoded BPSK symbol, and $\boldsymbol{z}[i]\in \mathcal{CN}(0,\boldsymbol{I})$ is the additive white Gaussian receiver noise assumed to be uncorrelated across the antenna array elements. The ambient signal is assumed to be complex Gaussian, i.e., $s[i]\in \mathcal{CN}(0,1)$, which is a practical assumption for OFDM systems \cite{Wei2010}. In this model, we assume that the ambient signal changes faster than the backscatter symbol does, and the ambient signal $s[i]$ is unknown at the AmBC receiver and cannot be decoded. In addition, the exact channel state information is not available at the AmBC receiver.

%

\section{Receiver operation} \label{sec:RxOperation}
\begin{figure}[!t]
\centering
	\includegraphics[width=0.9\columnwidth]{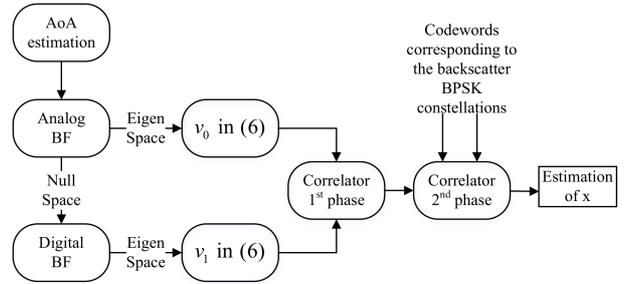}
	\caption{Receiver operation.}
    \label{fig:receiver_operation}
\end{figure}

The designed receiver aims to maximize the AmBC signal strength while nulling unwanted strong ones. The receiver seeks to find two beam-forming vectors $\boldsymbol{w}_l, \forall l=0,1$, as
\begin{equation*}
    \max_{\boldsymbol{w}_l} \quad \boldsymbol{w}_l^\dagger\boldsymbol{A}^{(0)}_l\boldsymbol{g}_l^{}\boldsymbol{g}_l^{\dagger}\boldsymbol{A}_l^{(0)\dagger}\boldsymbol{w}_l
\end{equation*}
subject to 
\begin{equation*}
    \boldsymbol{w}_l^\dagger\boldsymbol{C}_{l+1(\mathrm{mod~} 2)} = \boldsymbol{0}, \quad \text{and}\quad ||\boldsymbol{w}_l||^2=1,
\end{equation*}
where the constraint matrix is defined as
\begin{equation*}
    \boldsymbol{C}_l=\left[\begin{array}{ccc}\boldsymbol{A}_{l}^{(0)} & \boldsymbol{A}_l^{(1)} & \boldsymbol{A}_{l}^{(2)}\end{array}\right]
\end{equation*}
to null the unwanted signal in the presence of directional uncertainty. Fig.~\ref{fig:receiver_operation} depicts the operation process of the AmBC receiver. 

Define a Hermitian matrix $\boldsymbol{Q}_l=\boldsymbol{C}_l^{}\boldsymbol{C}_l^\dagger$ so that it can be diagonalized as   $\boldsymbol{Q}_l=\boldsymbol{U}_l\boldsymbol{\Lambda}_l\boldsymbol{U}_l^\dagger$, where $\boldsymbol{U}_l$ is an unitary matrix containing the eigenvectors and  $\boldsymbol{\Lambda}_l$ is a diagonal matrix containing the corresponding eigenvalues of  $\boldsymbol{Q}_l$. We assume that the number of receive antennas is larger than the number of multi-path components such that $\mathsf{rank}\{\boldsymbol{Q}_l\}\triangleq r_l<N$. Hence, we can take the first $|\mathcal{K}_l|$ eigenvalues to be non-zero and the rest $N-|\mathcal{K}_l|$ eigenvalues to be zero. We denote the eigenvectors, $\forall l=0, 1$, corresponding to the non-zero eigenvalues as $\boldsymbol{R}_l$ and the eigenvectors corresponding to the zero eigenvalues as $\boldsymbol{\Psi}_l$.

We consider a practical assumption in this paper shown in \eqref{eq:pathloss} that the physical channel in the scattered path suffers from much stronger loss than ordinary one-way direct channels. Hence, the receiver can initially perform AoA estimation for the strong direct paths using spectral-based or parametric approaches \cite{Krim1996}. Given a strong signal emanating from direct paths, the conventional beamformer (Bartlett) algorithm conducts AoA estimation, $\boldsymbol{\hat{\theta}}$, which maximizes the output power from certain direction as follows \cite{Krim1996}: 
\begin{equation} \label{eq:theta}
    \boldsymbol{\hat{\theta}}= \max_{\phi_k} \frac{\boldsymbol{a}(\boldsymbol{\Omega_k})^\dagger \boldsymbol{\hat{R}} \boldsymbol{a}(\boldsymbol{\Omega_k})}{\boldsymbol{a}(\boldsymbol{\Omega_k})^\dagger \boldsymbol{a}(\boldsymbol{\Omega_k})},
\end{equation}
where  $\boldsymbol{\hat{R}}$ is the sample co-variance matrix of the received signal sequence on the antenna array. The AoA estimation is conducted within one codeword.

\subsection{Two-stage beamformer}
Based on the estimated angles, the matrices $\boldsymbol{A}_0$, $\boldsymbol{R}_0$ and $\boldsymbol{\Psi}_0$ are calculated. The impact of the scattered paths on the direct path is considered to be negligible, and hence at the receiver we do not try to cancel them out. We propose to use a two-stage beamformer that in the first stage it maps the input vector to the eigenspace of the direct signal $\boldsymbol{u}[i]=\boldsymbol{U}_0^\dagger \boldsymbol{y}[i], \forall i=1,\cdots, M$. This vector can be further written as
\begin{IEEEeqnarray}{lll}
 \boldsymbol{u}_0[i] &=& \boldsymbol{R}_0^\dagger \boldsymbol{A}_0 \boldsymbol{g}_0s[i]+\boldsymbol{R}_0^\dagger\boldsymbol{A}_1 \boldsymbol{g}_1s[i]\tilde{x}[i]+\tilde{\boldsymbol{z}}_0[i] \nonumber\\
    &\approx& \boldsymbol{R}_0^\dagger \boldsymbol{A}_0 \boldsymbol{g}_0s[i]+\tilde{\boldsymbol{z}}_0[i], \label{eq:u0approx}
 \\
  \boldsymbol{u}_1[i] &=&\boldsymbol{\Psi}_0^\dagger\boldsymbol{A}_1 \boldsymbol{g}_1s[i]\tilde{x}[i] + \tilde{\boldsymbol{z}}_1[i].
  \label{eq:u1}
\end{IEEEeqnarray}
The approximation in (\ref{eq:u0approx}) is applied because $||\boldsymbol{g}_0||>>||\boldsymbol{g}_1||$ and after automatic gain control and analog-to-digital transmission, the scattered path vanish to the noise. The noise remain uncorrelated: $\tilde{\boldsymbol{z}}_0\sim\mathcal{CN}\left(\boldsymbol{0},\boldsymbol{I}_{r_0}\right)$ and $\tilde{\boldsymbol{z}}_1\sim\mathcal{CN}\left(\boldsymbol{0},\boldsymbol{I}_{N-r_0}\right)$.

In the second stage, we estimate the covariance matrices $\boldsymbol{\Sigma}_l=\mathbb{E}\left\{\boldsymbol{u}^{}_l\boldsymbol{u}_l^\dagger\right\}$ for $l=0,1$ and perform beamforming
\begin{equation*}
    \boldsymbol{v}_l = \mathsf{argmax}_{||\boldsymbol{v}_l||=1}\left\{\boldsymbol{v}_l^\dagger \boldsymbol{\Sigma}_l^{} \boldsymbol{v}_l^{}\right\}.
\end{equation*}
The solution coincides with the eigenvector of $\boldsymbol{\Sigma}_l$ that corresponds to the largest eigenvalue $\mu_l$.

Finally, we have two estimates $\nu_l=\boldsymbol{v}_l^\dagger \boldsymbol{u}_l^{}$, $l=0,1$,
\begin{equation}
\begin{split}
    \nu_0[i] &\approx h_0 s[i] +\zeta_0[i], \\ \nu_1[i] &= h_1 s[i] \tilde{x}[i] + \zeta_1[i], \quad \forall i=1,\cdots,M,
\end{split}
    \label{eqn:sig_beamformer}
\end{equation}
where $h_0=\boldsymbol{v}_0^\dagger\boldsymbol{R}_0^\dagger \boldsymbol{g_0}$, $h_1=\boldsymbol{v}_1^\dagger\boldsymbol{\Psi}_0^\dagger \boldsymbol{g_1}$, ${\zeta}_l[i]=\boldsymbol{v}_l^\dagger\tilde{\boldsymbol{z}}_l$, $l=0,1$. These two sequences in \eqref{eqn:sig_beamformer} will be applied to the proposed two-phase correlator presented in the following subsection.

\subsection{Two-phase Correlator}
Although the strong interference from the direct path has been greatly reduced shown in \eqref{eqn:sig_beamformer}, the unknown ambient signal remains in $\nu_1$ so that the receiver cannot detect $\tilde{x}[i]$. The first phase of the proposed correlator is to eliminate the unknown phase of $s$. We can study the cross-correlation of the above two measurements
\begin{equation}
   \tilde{\nu}[i] = \nu_0[i]^* \nu_1[i] = h_0^*h_1^{}|s[i]|^2 \tilde{x}[i] + \tilde{\zeta}[i],
\end{equation}
where $\tilde{\zeta}[i] = s[i]^{\dagger}h_0^{\dagger}\zeta_1[i] + \zeta_0[i]^{\dagger}h_1 s[i]\tilde{x}[i] + \zeta_0[i]^{\dagger}\zeta_1[i]$.

In the second phase of the correlator, firstly, the sequence  $\boldsymbol{\tilde{\nu}}=\{\tilde{\nu}[i]\},\;i=1,\cdots,M$ is correlated to the known two codewords associated to the backscatter BPSK symbols, i.e., $\boldsymbol{\tilde{\nu}}\boldsymbol{c}_l^T , \; l = \{1,2\}$; secondly, the squared absolute value of $\boldsymbol{\tilde{\nu}}\boldsymbol{c}_l^T$ is calculated. Then the decision is made by comparing the energy level. For codeword $l$, the output of the second phase of the correlator yields
\begin{equation}
   \Gamma_l = \left|\boldsymbol{\tilde{\nu}}\boldsymbol{c}_l^T \right|^2, \; \forall l = 1, 2.
\end{equation} 
The decision rule for the estimate of the AmBC BPSK signal becomes
\begin{equation}
    \setlength{\nulldelimiterspace}{0pt}
    \hat{x}=\left\{\begin{IEEEeqnarraybox}[\relax][c]{l's} +1, & for $\Gamma_1 \geq \Gamma_2$,\\
    -1, & otherwise.
    \end{IEEEeqnarraybox}\right.
\end{equation}

\paragraph*{Remark} The proposed receiver design can deal with the high power differences between the undecodable strong ambient signal and the weak backscatter information signal. In addition, the receiver does not need the channel state information which is challenging in practice to obtain because of  the nature of AmBC. Moreover, the receiver does not need to measure the power  of additive white Gaussian noise.  

\section{Numerical Results}\label{sec:Sim}
To validate the proposed design, we consider a practical setup such as the digital television broadcasting system that the ambient RF source is 1 km away from the AmBC receiver antenna array and the carrier frequency of the ambient system is 500 MHz. Recall that the sole frequency-dependent loss of the backscattering path can be approximately $-27$dB lower at 500MHz frequency band than that of the direct path. Without coding gain, a 30 dB received SNR per antenna of the direct link results in only a 3 dB SNR per antenna of the backscattered signal.

We consider that the linear receive antenna array has 8 elements, the AoA estimation is carried out using the Bartlett method, the ambient RF source is located at $60^\circ$, and the backscatter is located at $90^\circ$. The antenna separation of the receiver antenna array is set to half of the wavelength of the ambient signal, i.e., $\Delta d = \lambda/2$. In the results, the notation $H_{M}$ denotes a length-$M$ Hadamard codeword. Note that our proposed scheme only needs to measure the AoA, and does not decode the ambient signals and measure the noise power so that the receiver design can be dramatically simplified.

\begin{figure}[!t]
\centering
	\includegraphics[width= 0.9\columnwidth]{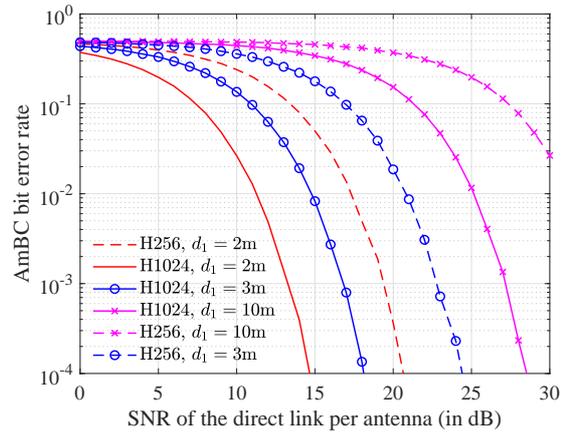}
	\caption{BER as a function of the received SNR of the direct link.}
    \label{fig:BER_Fc500M}
\end{figure}

Fig.~\ref{fig:BER_Fc500M} shows the BER of the AmBC system as a function of the received SNR per antenna of the direct link. The pathloss including frequency-dependent loss of the backscattered link is approximately 32 dB, 36 dB, and 46 dB higher than that of the direct link for different values of the distances between the backscatter and the AmBC receiver, i.e., $d_1=2m, 3m$, and $10m$, respectively. Such low SNR results in significant design challenges for AmBC receiver designs when the backscatter adopts BPSK or other higher-order modulation schemes even without interference. In order to operate the AmBC with an acceptable BER, the backscatter device is required to operate either within a short distance to the receiver, or applying a long codeword. For instance, the solid and the dashed curves show that when a length-1024 codeword is applied the backscatter can operate at a distance $d_1=2$ meters with a BER of $10^{-3}$ and the required SNR of the direct link per antenna is only around 13 dB. However, when $d_1=10$ meters, the required SNR has to be approximately 19.5 dB.

\begin{figure}[!t]
\centering
	\includegraphics[width= 0.9\columnwidth]{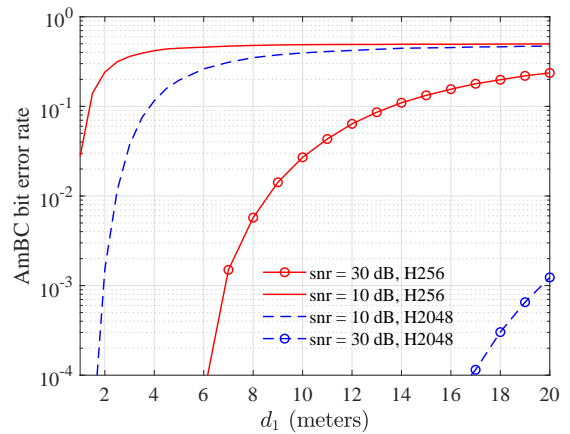}
	\caption{BER as a function of $d_1$.}
    \label{fig:BER_d1}
\end{figure}

Fig.~\ref{fig:BER_d1} depicts the BER as a function of $d_1$, the distance between the backscatter device and the AmBC receiver. The results indicate that our proposed scheme enables an AmBC device adopting a length-2048 codeword to operate with $10^{-3}$ BER at a distance of approximately 20 meters from the AmBC receiver when the received SNR of the legacy signal is 30 dB. If the SNR drops to 10 dB which in practice indicates a bad SNR for a practical broadcasting system, the proposed design allows the AmBC to operate with a $10^{-3}$ BER at a distance of approximately 2 meters from the AmBC receiver.

\section{Conclusions} \label{sec:Con}
In this paper, we have designed a two-stage hybrid beamformer for high dynamic range ambient backscatter receivers so that the challenges for receiver design can be reduced. The design allows the receiver automatic gain control to adjust to the level of the weak scattered paths so that common analog-to-digital converters can be applied for sampling the signal. In addition, the proposed novel design allows the AmBC receiver to detect the backscatter BPSK signals without decoding the ambient RF signals and requiring knowledge of the instantaneous channel state information. The results have confirmed that the proposed design enables an AmBC device to operate at a reasonable distance from the AmBC receiver with $10^{-3}$ BER without applying error correction techniques.


\section*{Acknowledgment}
This work was supported in part by the Academy of Finland under Project No. 311760 and Project No. 319003, and the U.S. NSF under Grant CNS-1702850.
%

\ifCLASSOPTIONcaptionsoff
  \newpage
\fi

\bibliographystyle{IEEEtran}
\IEEEtriggeratref{24}

\end{document}